\documentclass{article}
\usepackage{setspace}
\usepackage[english]{babel}

\usepackage[colorlinks=true, allcolors=blue]{hyperref}
\usepackage[a4paper,top=2cm,bottom=2cm,left=3cm,right=3cm,marginparwidth=1.75cm]{geometry}
\usepackage{enumitem} 
\usepackage{amsmath} 
\usepackage{amssymb} 
\usepackage{bm} 
\usepackage{tikz} 
\usepackage{mathtools} 
\usepackage{graphicx} 
\usepackage{harvard} 
\usepackage{titlepic} 
\usepackage[export]{adjustbox} 
\usepackage[ruled,vlined]{algorithm2e} 
\usepackage{pdfpages}
\usepackage{caption}
\usepackage{soul} 

\graphicspath{ {./images/} }

\title{Forecasting overhead distribution line failures using weather data and gradient-boosted location, scale, and shape models.}
\author{Antoni Michał Siemiński}

\begin{document}

\renewcommand{\abstractname}{Abstract}
\Large
 \begin{center}
Forecasting overhead distribution line failures using weather data and gradient-boosted location, scale, and shape models.\\ 

\hspace{10pt}

\large
Antoni M. Sieminski$^1$, Dr Carl R. Donovan$^2$ \\

\hspace{10pt}

\small  
$^1$) University of Edinburgh \\
s2410784@ed.ac.uk\\
$^2$) University of St Andrews \\
crd2@st-andrews.ac.uk 

\end{center}

\hspace{10pt}

\normalsize

Overhead distribution lines play a vital role in distributing electricity, however, their freestanding nature makes them vulnerable to extreme weather conditions and resultant disruption of supply. The current UK regulation of power networks means preemptive mitigation of disruptions avoids financial penalties for distribution companies, making accurate fault predictions of direct financial importance. Here we present predictive models developed for a UK network based on gradient-boosted location, scale, and shape models, providing spatio-temporal predictions of faults based on forecast weather conditions. The models presented are based on  (a) tree base learners or (b) penalised smooth and linear base learners -- leading to a Generalised Additive Model (GAM) structure, with the latter category of models providing best performance in terms of out-of-sample log-likelihood. The models are fitted to fifteen years of fault and weather data and are shown to provide good accuracy over multi-day forecast windows, giving tangible support to power restoration.

\section{Introduction}

Reliable electricity supply has become essential in all industrialised countries. Power outages are more than an inconvenience: most industries cannot operate without electricity, which means that unscheduled power outages have a severe impact on the economy \cite{Poudineh2017ElectricityEconomy} and can even lead to death. The strategic importance of power supply cannot be overstated, hence most governments choose to oversee this sector. In the UK, the power distribution industry is a regulated monopoly, meaning infrastructure is provided and maintained by private companies, but their performance is independently assessed by government regulators leading to financial penalties if expected standards are not met. Case in point, excessive power outages experienced under reasonably predictable conditions can lead to fines. This creates an important incentive for companies to invest in predictive modelling to aid quick energy restoration, for example by preemptive distribution of repair resources.

Most of the energy supply chain uses overhead distribution lines which are cheaper, but also more vulnerable, than underground lines \cite{Maliszewski2012EnvironmentalArizona}. These structures are constantly exposed to changing environmental conditions, which are often responsible for the faults \cite{Mukherjee2018AU.S.}. For example, an iced overhead line can start oscillating with the wind, gain momentum, and snap. Other factors include flocks of birds flying into the lines, or high soil moisture leaving vegetation more vulnerable to falling onto the lines in high wind conditions. Power outages can also be caused by non-weather events  \citeaffixed{Mukherjee2018AU.S.}{e.g., intentional cyber and physical attacks, equipment failure or fuel supply emergency;}, but their effects are much less important than those of extreme weather  -- as measured by the outage duration and the number of customers affected \cite{Mukherjee2018AU.S.}. The main aim of this paper is to create a state-of-the-art model, which will predict the probability distribution of the number of faults given weather forecasts. 

\subsection{Previous models}

So far, most of the research in this area was based in the US and often focused on only a small subset of the environmental conditions (mostly hurricanes), while neglecting others, such as thunderstorms \cite{Kabir2019}. Fortunately, their statistical methodology also applies to other weather-induced faults. Other shortcomings include the failure to account for the characteristic zero-heavy distributions of fault counts, which can lead to inaccurate predictions and underestimation of predictors' effects. Furthermore, all datasets are spatio-temporal even if not explicitly modelled as such, which opens the problem to many more sophisticated models.

\subsubsection{Parametric models}

Early attempts at fault prediction were based on (semi-) parametric statistical models, such as Generalised Linear Models (GLMs) \cite{Han2009ImprovingModels} and their extensions for (a) nonlinear relationships (Generalised Additive Models; GAMs) \cite{Han2009ImprovingModels} and/or (b) nonindependent errors (Generalised Linear Mixed Models; GLMMs or Generalised Additive Mixed Models; GAMMs) \cite{Liu2008SpatialStorms}. These models explicitly predicted the full outcome distribution, which allowed for probabilistic evaluation of the uncertainty around the individual observations. Given a weather forecast, the decision makers were able to assess the likelihood of fault counts and make an informed decision based on the data. 

However, in these analyses, the distributions did not account for the ``inflation" of zero counts and temporal autocorrelation of errors, a violation of assumptions that questions inferential validity. Certainly common model selection procedures are sensitive to these properties, typically leading to overstated covariate effects \cite{Tibshirani2016}.

Other statistical approaches involved predicting the distribution of outage duration after major storms using accelerated failure models within the GAM framework \cite{Liu2007StatisticalStorms}. This approach could also be used to predict individualised times to overhead distribution lines' failures, thus allowing for targeted maintenance scheduling. However, this approach can rarely be implemented as it would require highly precise weather forecasts and records on every overhead distribution line.

\subsubsection{Machine Learning (ML) approaches}

Methods typically under the ML banner have quickly risen in popularity and have ended up dominating the field \cite{Kabir2019}. These methods are characterised by less effort in both the preprocessing of the data and model specification as much of it is often handled internally within the model fitting procedure. The relationships between responses and covariates are rarely known by the analyst in advance, so these highly data-driven methods usually lead to better predictive performance than parametric and semi-parametric models. 

\paragraph{Single-stage ML}

\citeasnoun{Guikema2010} compared predictive performance of (semi-) parametric models (GAMs and GLMs) and non-parametric tree-based models to predict number of faults. For their data, the latter group of models clearly outperformed the former. Although, it is common knowledge that these non-parametric models are advantageous in terms of prediction, it is important to acknowledge the ways in which the comparison was flawed. The outcomes were compared using Mean Absolute Error (MAE) as their performance metric, which favours good models of the median and neglects accurate prediction of the error distribution - inference being typically, or necessarily, ignored when employing ML. For example, GLMs and GAMs had the advantage of having easily accessible zero-inflated distributions \citeaffixed{Zuur2009}{e.g.,}, but this advantage was not utilised in the comparison. Notably prediction intervals are of clear utility in many predictive modelling situations, but typically ignored in the deployment or selection of ML models - characterisation of error distributions being essential to achieve these.

More complex applications of ML included: Artificial Immune Recognition Systems \citeaffixed{Xu2006UsingData}{AIRS;} and E-algorithms \cite{Xu2007PowerAlgorithm} for multiclass classification of outage causes. \citeasnoun{Silva2020TowardsSalford} extended these approaches by using Artificial Neural Networks to predict additional outage characteristics, such as outage duration. \citeasnoun{Zhou2007ModelingLines} used a probabilistic (Bayesian) network model to predict the number of faults with the mean of the Poisson distribution. This approach was one of the few which provided prediction intervals within a machine learning framework -- although, it still used a simplistic non--zero-inflated distribution. Bayesian models, or statistical approaches generally, further enhance the results by providing uncertainty around individual parameter values, such as the mean. This is also the case for the Bayesian Adaptive Regression Trees \citeaffixed{Guikema2010}{e.g., used in}, but this feature has not been utilised in the literature. 

\paragraph{Two-stage modelling}

Nevertheless, many authors acknowledged the problem of zero-inflation and tried to overcome it with a two-stage modelling approach \cite{Guikema2012HybridData,Kabir2019,McRoberts2018,Quiring2014IncorporatingModeling}, separately modelling the probability of a zero count, then the distribution of counts conditional on being non-zero. \citeasnoun{Guikema2012HybridData} and \citeasnoun{Quiring2014IncorporatingModeling} used a classification tree model \cite{Breiman1984} for the first step and a Poisson GAM for the second, with \citeasnoun{McRoberts2018} showing improved point estimates by changing both models to a Random Forest (RF) \cite{Breiman2001,Friedman2000} .

The two-stage models focused on delivering accurate point estimates but neglected the uncertainty around specific predictions. \citeasnoun{Kabir2019} extended the method by using probability predictions to \textit{simulate} labels for the first step, and then used these labels to train two quantile regression models for each of the simulated labels.

The general problem with this two-stage approach, however, is its reliance on classifying observations instead predicting their probability. The distribution of faults was strongly zero-inflated, which led to the models predicting the zero-count category more often because of its prevalence rather than the predictors' values - effectively leading to strongly imbalanced classes. Typical mitigations are through increased cost of misclassified minority classes by weighting, upsampling, or downsampling \citeaffixed{Elkan2002}{cost-sensitive learning; see}. Each method works well when the misclassification costs are known in advance and fixed over time yet this is may be difficult to set for real-life scenarios.

\subsection{Problem specification}

Broadly, this is a regression problem. The data are described in detail in section \ref{sec:data}, but a high-level description follows to inform our model selection. 

Fault data is collected continuously by distribution companies, logging the time of detection on the network, the rough spatial location (known immediately to a substation level - although exactly where on a line is not) and eventually a purported cause. This provides a spatio-temporal dataset of faults, which in our case is aggregated to management region and day resolution i.e. the number of faults per day within each region. Observed weather data is collected on fine spatio-temporal scale, as well as regional weather forecasts several days into the future, variously at 6-hour to 1-day scales. The problem, therefore, is to predict the number of regional weather-related faults per day, at various time-scales into the future, on the basis of associated weather forecasts. Distribution networks are by necessity robust, meaning faults are rare, and the data, therefore, heavy in zeros. In this particular case, the power distributor's spatial management regions do not fully match the regions that fault counts are collected for - consequently a weighted allocation of faults is performed, meaning the response is non-negative continuous, rather than strictly count.

Location, scale, and shape (LSS) models seem like a natural choice for this scenario as they offer considerable flexibility in fitting complex error distributions (including zero-heavy distributions) and signal.
We have chosen to train a variety of gradient-boosted models with either (a) GAM smoothers or (b) the regression tree base learners \cite{Hofner2016GamboostLSS:Framework,Thomas2018GradientUpdates}. The mixing parameter of the zero-adjusted distribution was estimated with a separate probabilistic classifier using either (a) a gradient-boosted GAM \cite{Hothorn2005,Hofner2014,Hothorn2021Package} or (b) a gradient-boosted tree model known as XGBoost  \cite{Chen2016Xgboost:Systemb}.

The problems of collinearity, concurvity, and model selection were all automatically handled by gradient boosting \cite{Hofner2011ABoosting}. The  hyperparameters were selected by using a holdout dataset. The LSS models and the GAM classifier were boosted until the out-of-sample predictions stopped improving. The remaining hyperparameters for the tree models were searched with the:
\begin{itemize}
    \item Genetic Algorithm  \cite{Whitley1994ATutorial} for the gradient boosted trees for Location, Scale, and Shape 
    \item Bayesian Optimisation \cite{Snoek2012PracticalAlgorithms.} for the XGBoost probabilistic classifier
\end{itemize}  

The final model performance was estimated using another holdout dataset in order to estimate the extent of overfitting due to hyperparameter search. This section provides a short theoretical background on the \hyperref[sec:methods]{methods} used for the project.

\paragraph{Zero-heavy distributions} \label{sec:zadj}

Another common mixture of distributions are called zero-inflated, zero-heavy or zero-adjusted. Such distributions have an additional mixing parameter $\xi_0$, which denotes the probability that a zero-count is observed, and its compliment is the probability of the observation coming from another distribution. According to \citeasnoun{Stasinopoulos2017ZeroLine}, the zero-adjusted distributions are nonnegative continuous, whereas zero-inflated distributions are nonegative integer distributions. Because our data consisted of weighted averages of counts, the zero-adjusted distribution was most appropriate, and its general form is presented in equation \ref{eq:zadj}. 

\begin{align} \label{eq:zadj}
    f_Y(y| \bm{\theta}, \xi_0) = \begin{cases}
    \xi_0, & \text{if $y = 0$}.\\
    (1 - \xi_0)f_W(y | \bm{\theta}), & \text{if $0 < y < \infty$}
  \end{cases}
\end{align}

In equation \ref{eq:zadj}, $f_Y(y | \bm{\theta}, \xi_0)$ is the probability density function in the range of $[0; \infty)$, $f_W$ is the probability density function of a continuous distribution in the range of $(0; \infty)$ with the parameter vector $\bm{\theta}$; $\xi_0$ is the mixing parameter and the probability mass for all observations equal to zero.

Zero-adjusted distributions can offer a significant computational advantage over alternative distributions with the same number of parameters. As shown in equation \ref{eq:zadj_proof}, they allow the model to be fitted in two independent steps: (a) classifying all data as either zero or nonzero and (b) predicting the number of nonzero observations after removing all zero observations from the training data \citeaffixed{Stasinopoulos2017ZeroLine}{as implemented in the \texttt{gamlss.inf} package;}. The two-model method -- for zero-adjusted distributions -- is derived in equation \ref{eq:zadj_proof}.

\begin{equation} \label{eq:zadj_proof}
\begin{split}
    \mathcal{L}_Y(\bm{y} | \bm{\theta}, \xi_0) & =  \prod_{i = 1}^{N} f_Y(y_i| \bm{\theta}, \xi_0) \\
    & = \prod_{i = 1}^{N} \xi_0^{I(y_i = 0)} 
            \Big\{
                (1 - \xi_0) f_W(y_i | \bm{\theta})
            \Big\}^{I(y_i > 0)} \\
    & = \prod_{i = 1}^{N} \xi_0^{I(y_i = 0)}
                (1 - \xi_0)^{I(y_i > 0)}
                f_W(y_i | \bm{\theta})^{I(y_i > 0)} \\
    & = 
    \underbrace{ 
        \prod_{i = 1}^{N} \xi_0^{I(y_i = 0)}(1 - \xi_0)^{I(y_i > 0)}}_{\mathclap{\substack{\text{Bernoulli likelihood} \\ \text{for all observations}}}}
    \quad \times
    \underbrace{ 
        \prod_{j = 1, y_j > 0}^{N}f_W(y_j | \bm{\theta})}_{\substack{
        \text{Continuous distribution's likelihood} \\
        \text{for all nonzero observations}}} \\
    & = \Pr(Y = 0|\xi_0) \times \mathcal{L}_W(Y|Y > 0, \bm{\theta})
\end{split}
\end{equation}

This allows the model to be fitted in two completely independent stages, while preserving its likelihood interpretation: 
\begin{itemize}
    \item Fitting a binary Bernoulli model to all data to predict whether an observation is equal to zero
    \item Fitting a model with a positive continuous distribution by using all nonzero observations
\end{itemize}
The overall likelihood of the final model will be a Bernoulli likelihood (for whether the observation is equal to zero) times the likelihood of the chosen continuous distribution for all nonnegative observations. 

For overhead distribution lines, the majority of observations have the response equal to zero, so the two-stage method leads to fast model convergence. This has no effect on the asymptotic properties of time complexity -- in terms of the big \textit{O} notation -- but is highly relevant regardless of it. Furthermore, additional gains come from factoring $\xi_0$ out of the parameters of the distribution as it reduces the number of simultaneously modelled parameters leading to a faster fit -- again, with no change to the big \textit{O}. Lastly and most importantly, this split allows for more flexibility in terms of:

\begin{itemize}
    \item distribution choice \citeaffixed{Rigby2021DistributionsShape}{\texttt{gamlss.dist} has many continuous positive distributions available, but only two of them have their zero-adjusted version}
    \item model choice (there are many widely available methods for modelling $\xi_0$ but few for predicting LSS distributions)
\end{itemize}

\paragraph{Flexible models for the systematic component} \label{sec:model_selection}

The relationship between weather and faults is likely to be complex and not well known \textit{a priori}, meaning data-driven forms with little assumption are preferred - interactions being particularly difficult to specify. Boosted methods are particularly popular within ML for approximating functional relationships between response and covariates with little careful specification \cite{Mayr2012TheBoosting}. The original boosting algorithm was based on iteratively refitting simple decision trees (stumps) and increasing the weights of misclassified observations after each iteration \cite{Freund1996ExperimentsAlgorithm}. The final model was a weighted sum of all trees (hence an ensemble method) with higher weights for better-performing trees as measured by the negative log odds of a pseudo-loss function. Gradient boosting is more general than that and can be used to produce models of similar form to the ones generated by the previously described shrinkage methods \cite{Vogt2020OnLasso}.

Model-based boosting (for LSS) works by iteratively refitting base learners to each parameter and selecting the one which maximises the overall log-likelihood of the model. A conceptual version of \possessivecite{Thomas2018GradientUpdates} algorithm from \texttt{gamboostLSS} package is given in algorithm \ref{alg:noncyclical_boosting}.

\begin{algorithm}[!ht]
\DontPrintSemicolon
\caption{Noncyclical Gradient Boosting for Location, Scale, and Shape\label{alg:noncyclical_boosting}}
\textbf{Inputs:}\;
\begin{itemize}[noitemsep,topsep=0pt]
    \item Distribution $\mathcal{D}(\bm{\theta})$ with parameters $\theta_k$, where $k = 1,\dots, K$
    \item Link functions $g_k(.)$ for each parameter $\theta_k$
    \item Base learners $h_{1_k}(\bm{X}^{(1_k)}),\dots,h_{p_k}(\bm{X}^{(p_k)})$ for each parameter $\theta_k$
    \item Number of boosting iterations $m_{stop}$ and shrinkage parameter $\nu$
    \item Data with columns of appropriate class (i.e. \texttt{factor}, \texttt{ordinal}, and/or \texttt{numeric})
    \item Optionally, additional data for specific base learners (e.g., smoothing matrices, custom bases, boundary knots)
\end{itemize}
\textbf{Initialise:}\;
\begin{itemize}[noitemsep,topsep=0pt]
    \item Find offset values $\hat{\eta}_{\theta_k}^{[0]}$ for each parameter from $k$
\end{itemize}
\textbf{Iterate:}\;
\For{$m$ in $\{1,\dots,m_{stop}\}$}{
 \For{$k$ in  $\{1,\dots,K\}$}{
  Compute the gradient of the log-likelihood function $\ell_{\mathcal{D}}(y, \bm{\theta}^{[m-1]})$ for each response $y$ with respect to the linear predictor of $\theta_k$ at the $m-1$'th iteration  $\hat{\eta}_{\theta_k}^{[m-1]}$
  
  $\bm{u}_k^{[m]} \gets \frac{\partial }{\partial \hat{\eta}_{\theta_k}^{[m-1]}} \ell_{\mathcal{D}}(\bm{y}, \bm{\theta}^{[m-1]})$ \;
  \For{$i$ in $1,\dots,p_k$}{
   Fit the base learner $h_{i_k}(\bm{X}^{(i_k)})$ to the vector $\bm{u}_k^{[m]}$ to minimise the residual sum of squares\;
  } \;
  Save the base learner with the lowest residual sum of squares\;
 }\;
 Out of the K base learners for each of the parameters $\theta_k$, shrink the influence of each by the shrinkage parameter $\nu$.\;
 Then, add the one which maximises the overall likelihood by adding it to its linear predictor $\hat{\eta}_k$ for the parameter $k$\;
}\end{algorithm}

Base learners can be of various forms and include decision trees, penalised splines, and Gaussian Markov Random Fields, but are typically chosen to be simple - complexity where needed is achieved by combining many such simple building blocks. Gradient boosting is also known as functional gradient descent as, similar to gradient descent, it is an iterative optimisation routine based on taking small steps to reach the function's optimum \cite{Mason1999BoostingSpace}.
The step size parameter $\nu$ is analogous to the step size in gradient descent, and it needs to be sufficiently small to prevent the routine from overshooting the optimum. Contrary to gradient descent, gradient boosting is rarely run with the sole purpose of maximising the within-sample performance. Iterations are purposefully stopped early using the parameter $m_{stop}$ in order to reduce the model's variance at a small expense of increased bias.

The basis functions employed here included B-Splines with knots placed at equidistant intervals and the simple dummy-coded matrix. The former was used with a P-Spline penalty, which penalises the differences between adjacent bases' coefficients \cite{Eilers1996FlexiblePenalties}. The latter was either used with either a simple ridge penalty or with a spatial penalty matrix reducing the differences between adjacent regions' coefficients for spatial effects (Markov Random Field).

\subsection{Hyperparameter tuning} \label{sec:hyperparameter_tuning}

The hyperparameters for the tree models were tuned with Gaussian Process optimisation or the genetic algorithm with the goal of maximising the out-of-sample log likelihood on a holdout dataset. The GAM models used only one hyperparameter -- the number of iterations $m_{stop}$ -- and optimised it exactly by boosting until no improvements were observed in the last 100 iterations.

\section{Data}
\label{sec:data}

All observations came from Scotland, which has a mild temperate oceanic climate resulting in low temperature amplitudes and few thunderstorms throughout the year. Southern Energy Power Distribution (part of SSE plc group) has a monopoly over managing overhead distribution lines in Scotland, and its database was used to obtain regional fault counts from October 2015 until 2019. 

The weather data were provided by the MeteoGroup, which offered forecasts for up to five days in advance. Each day was measured as starting and ending at 6am. Weather forecasts were provided for slightly different regions than the company's administrative regions, for which the fault counts were recorded. This resulted in the necessity to use a weighted average of these counts based on the percentage of the overlapping area between the regions.

There were two kinds of forecasts: (a) one for the weather of the upcoming day and the next day (short forecast), and (b) a different and less precise one for the next three consecutive days (long forecast). Here we focus on short forecast data for nine Scottish regions: Perth \& Angus, Aberdeenshire, Moray Firth, Argyll \& West Highland, North West Highland \& Skye, Central Highlands, Western Isles, Orkney \& North East Caithness, and Shetland. These offer projections of the following weather conditions:
\begin{itemize}
    \item risk (indicated by a three-level colour system: green, yellow, and red)
    \item temperature on the day (minimum and maximum values)
    \item wind forecasts for four six-hour intervals (included: direction, gust, and mean speed)
    \item rain for two twelve-hour intervals (minimum and maximum values in mm)
    \item snow for two twelve-hour intervals (depth, icing, and height)
    \item lightning category for two twelve-hour intervals (using an ordinal scale from one to five)
\end{itemize}

Overall, some 12 covariates were available for fault prediction.

\section{Methods} \label{sec:methods}

\subsection{Models for the systematic component}

As described in section \ref{sec:zadj}, each dataset would benefit from splitting its modelling into two independent stages: (a) fitting a binary model for the detection of zero counts and (b) fitting a positive continuous model for the number of faults. For both cases, a tree-based and a GAM-based model were trialled to see which one performed better. All continuous terms were centred and scaled before modelling for computational reasons. 

\subsubsection{GAM formulation}

All GAM-based models had a similar formula specification, which was done by decomposing different model terms. The model included a separate linear intercept base learner, and all other learners had their intercepts omitted. Icing was a binary variable, which was modelled with a simple dummy-coded linear learner. Wind direction was binned into eight categories and modelled as a categorical covariate with an $L_2$ penalty forcing its degrees of freedom $df = tr(2\bm{S} - \bm{S}^\intercal\bm{S})$ to one. This form was the result of software and methodological limitations as \possessivecite{Fahrmeir2004PenalizedPerspective} method implemented in \texttt{mboost} does not work for cyclic smooths. Lightning category and risk were both modelled as ordinal covariates with the same $df$ specification and a second-order difference penalty on the adjacent levels of their categories. All continuous variables were decomposed into a linear and a smooth term (with $df = 1$), and these included: \textit{maximum wind gust}, \textit{mean wind}, \textit{minimum temperature}, \textit{maximum temperature}, \textit{minimum rain}, \textit{maximum rain}. The regional effects were modelled as an $L_2$ penalised categorical variable with $df = 1$. The dataset with one- to two-day forecasts also had a spatial random effect based on a decomposed (i.e. centred) Gaussian Markov Random Field smoother.

Additionally, four of these variables were also decomposed to build two three-way interactions between wind variables (gust or mean) $\times$ region $\times$ wind direction, which naturally led to the inclusion of all second-order interactions. Because region and wind direction were coded as categorical variables, there were no problems with the model applying an isotropic smooth penalty to terms with different units. The variance parameter $\sigma$, the optional kurtosis parameter $\tau$, and the mixing parameter $\xi_0$ were also predicted with different intercepts for one- and two-day forecasts. The rest of the formula was the same for all parameters.

\subsubsection{Tree formulation}

The tree models were also fitted to the same data as GAM models and used a single tree base learner with all variables. The only difference was that \textit{wind direction} was used as a continuous variable instead of categorical. 

The binary model was fitted with \texttt{xgboost} and \texttt{tidymodels} packages and required manual dummy-coding of categorical variables and integer coding of continuous variables. The positive-continuous model was fitted using the \texttt{gamboostLSS} package, which uses the \texttt{partykit} package and the \texttt{ctree} function as its back-end. All categorical and ordinal variables were modelled using the default settings, that is dummy coding for the categorical variables, and distance coding for ordinal variables \citeaffixed{Hothorn2015Ctree:Trees}{see}.

\subsubsection{Hyperparameter tuning details} \label{sec:hyper_details}

All models except for XGBoost used the \texttt{mboost} package as their back-end, which allowed for adaptive increases in models' iterations. A \texttt{while} loop was written to iteratively increase the number of iterations by 50 until no likelihood improvement was observed on the holdout dataset between the two loop iterations. This avoided the need for optimising this hyperparameter by trial and error. The learning rate for GAM-based models was set to $0.3$, because all terms were limited to just one degree of freedom. 

As for the tree-based models, XGBoost was trained first due to the ease with which the \texttt{tidymodels} package allows for automatic hyperparameter tuning via Bayesian optimisation. The optimised hyperparameters included: the number of iterations $m_{stop}$, the learning rate $\nu$, the proportion of sampled rows on each iteration, the proportion of sampled columns on each iteration, the minimum number of observations on each terminal leaf, the maximum tree depth (where the root node is at zero depth), and the required loss reduction for a tree split. 

The tree-based LSS model had fewer parameters to optimise, but allowed for adaptive increases in the number of iterations. The minimum number of observations for a tree split was set to 200, the maximum number of interactions to three, and the minimum complexity reduction was set to zero. The maximum depth was set to be between one and four, the number of sampled columns to be between one and eight, and the minimum number of observations on a leaf to be one of: 50, 250, 450, and 650. This discrete specification was required as a binary version of the genetic algorithm was used. The algorithm used the following settings: $n_{population} = 8$, $n_{generations} = 3$, $p_{crossover} = 0.8$, $p_{mutation} = 0.1$, and $Elitism = 1$.

\subsubsection{Models for the stochastic component}

A simple Bernoulli distribution was used for the binary classification stage; however, for the positive continuous prediction, a wide variety of distributions were available for modelling \cite{Stasinopoulos2017ZeroLine}. Because the GAM model required less tuning than the tree models, it was used first to explore which distributions are worth using for future models. Each additional distribution parameter was likely to result in a much longer computing time, so a mix four distributions was trialled to begin with and included: Gamma (2 parameters), Generalised Gamma (3 parameters), Inverse Gaussian (3 parameters), and Box-Cox t-distribution (4 parameters, see equation \ref{eq:bct}) \cite{Rigby2006UsingKurtosis}. All distributions came from the \texttt{gamlss.dist} package \cite{Rigby2020DistributionsR} and used the default link functions from the \texttt{GA()}, \texttt{GG()}, \texttt{GIG()}, and \texttt{BCTo()} functions. 

\begin{align} 
\begin{split}
\label{eq:bct}
   f_{BCt}(y|\mu,\sigma,\nu,\tau) &=
\Gamma(\frac{\tau+1}{2})\times\Gamma(\frac{1}{2})\times \Gamma(\frac{\tau}{2})\times\tau^{\frac{1}{2}}\times(1+z^{(2/\tau)})^{-(\tau+1)/2}\times(y\sigma)^{-1} \\
z &= \begin{cases}
    (\frac{y}{\mu})^{\nu}-(\nu\sigma)^{-1}, \text{if } \nu \neq 0.\\
    log(\frac{y}{\mu})/\sigma, \text{otherwise} 
    \end{cases}
\end{split}
\end{align}

The four-parameter distribution outperformed all of its competitors, so the Box-Cox Power Exponential distribution using \texttt{BCPEo()} \cite{Voudouris2012ModellingGAMLSS} and the Generalised Beta 2 distribution using \texttt{GB2()} were also trialled. Negative gradients of the loss function were stabilised using median absolute deviation method from the \texttt{as.families()} function from the \texttt{gamboostLSS} package \cite{Fenske2021Package}, which means that each gradient was divided by its median absolute deviation after every boosting iteration. This ensured that all gradients had relatively similar ranges \cite[Appendix B]{Hofner2016GamboostLSS:Framework} and mitigated against non-convergence.

\section{Results}

\begin{figure}[h]
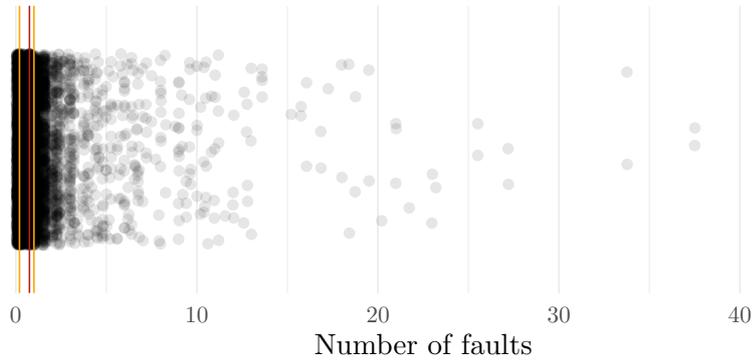

    \centering
    \include{figures/dist_nonzero_faults}
    \caption{Distribution of the number of faults after excluding zero counts. The red line indicates the median and the orange lines indicate the 50\% interquartile range.}
    \label{fig:dist_nonzero_faults}
\end{figure}

The majority of observations were equal to zero ($\approx 62\%$), and the remainder was highly left-skewed (see figure \ref{fig:dist_nonzero_faults}). 

\begin{figure}[ht]
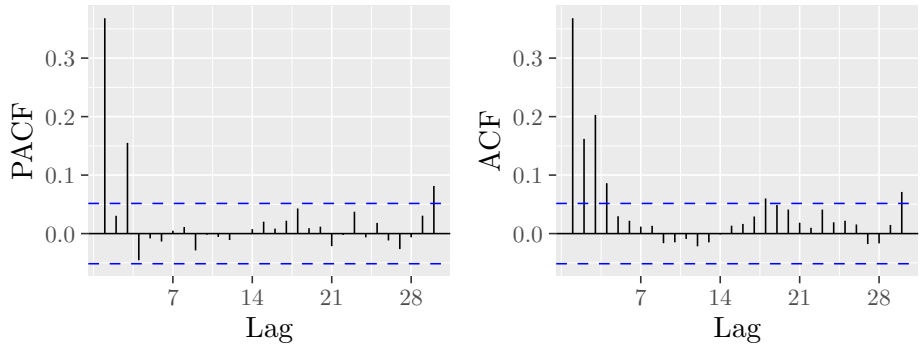

    \centering
    \include{figures/ts_plots}
    \caption{Autocorrelation and partial autocorrelation functions for the number of faults within each region. The dashed lines indicate a 95\% confidence interval.}
    \label{fig:acfs}
\end{figure}

Raw faults were significantly autocorrelated with a lag of four days, but after conditioning on other lagged observations, only lags one, three, and thirty were statistically significant (see figure \ref{fig:acfs}). The oscillatory behaviour of the raw autocorrelation suggests that it might stem from weather cycles, which can be accounted for by using weather forecasts.

The number of faults remained reasonably stationary throughout the years, albeit with slight seasonal patterns. Extreme weather conditions were associated with a higher number of faults.

\subsection{Hyperparameter tuning} \label{sec:res_short_forecasting}

The boosted GAM model performed slightly better than the XGBoost model for the classification of zero counts (see table \ref{table:bern}), so it was selected for refitting on all training data. The XGBoost tested 76 hyperparameter sets through Bayesian optimisation (with 15 starting models), whereas the GAMboost model was run only once until convergence.

\begin{table}
\centering
\begin{tabular}{ l*{4}{c} } 
\hline
Model & $\nu$ & $\ell_{total}$ & $m_{stop}$ & $\ell_{average}$ \\
\hline 
GAMboost     & 0.30 &  -2967 & 4098 & -0.565 \\ 
XGboost      & 0.01 &  -2976 & 1351 & -0.567 \\ 
\hline
\end{tabular}
\caption{Bernoulli models' out-of-sample performances for the step size parameter $\nu$ and stopping iteration $m_{stop}$}
\label{table:bern}
\end{table}

After selecting the logistic GAMboost model, the model was retrained on all data -- including the holdout data, and the results were interpreted. The final model had an in-bag $\ell_{total}\approx-11595$ and $\ell_{avg} \approx-.56$.

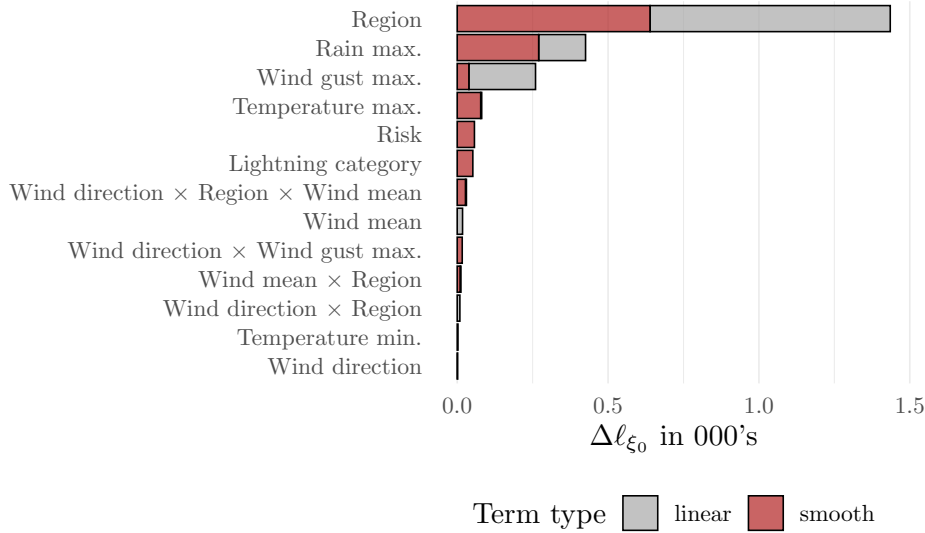
\begin{figure}[h]
    \centering
\begin{tikzpicture}[x=1pt,y=1pt]
\definecolor{fillColor}{RGB}{255,255,255}
\path[use as bounding box,fill=fillColor,fill opacity=0.00] (0,0) rectangle (361.35,216.81);
\begin{scope}
\path[clip] (177.52, 67.14) rectangle (355.85,211.31);
\definecolor{drawColor}{gray}{0.92}

\path[draw=drawColor,line width= 0.3pt,line join=round] (213.86, 67.14) --
	(213.86,211.31);

\path[draw=drawColor,line width= 0.3pt,line join=round] (270.34, 67.14) --
	(270.34,211.31);

\path[draw=drawColor,line width= 0.3pt,line join=round] (326.81, 67.14) --
	(326.81,211.31);

\path[draw=drawColor,line width= 0.6pt,line join=round] (185.62, 67.14) --
	(185.62,211.31);

\path[draw=drawColor,line width= 0.6pt,line join=round] (242.10, 67.14) --
	(242.10,211.31);

\path[draw=drawColor,line width= 0.6pt,line join=round] (298.58, 67.14) --
	(298.58,211.31);

\path[draw=drawColor,line width= 0.6pt,line join=round] (355.05, 67.14) --
	(355.05,211.31);
\definecolor{drawColor}{RGB}{0,0,0}
\definecolor{fillColor}{RGB}{169,169,169}

\path[draw=drawColor,line width= 0.6pt,line cap=rect,fill=fillColor,fill opacity=0.70] (185.62, 68.78) rectangle (185.79, 78.61);
\definecolor{fillColor}{RGB}{178,34,34}

\path[draw=drawColor,line width= 0.6pt,line cap=rect,fill=fillColor,fill opacity=0.70] (185.62,156.15) rectangle (192.07,165.98);

\path[draw=drawColor,line width= 0.6pt,line cap=rect,fill=fillColor,fill opacity=0.70] (185.62,145.23) rectangle (191.47,155.06);
\definecolor{fillColor}{RGB}{169,169,169}

\path[draw=drawColor,line width= 0.6pt,line cap=rect,fill=fillColor,fill opacity=0.70] (190.00,178.00) rectangle (214.94,187.83);
\definecolor{fillColor}{RGB}{178,34,34}

\path[draw=drawColor,line width= 0.6pt,line cap=rect,fill=fillColor,fill opacity=0.70] (185.62,178.00) rectangle (190.00,187.83);
\definecolor{fillColor}{RGB}{169,169,169}

\path[draw=drawColor,line width= 0.6pt,line cap=rect,fill=fillColor,fill opacity=0.70] (185.62,123.39) rectangle (187.62,133.22);
\definecolor{fillColor}{RGB}{178,34,34}

\path[draw=drawColor,line width= 0.6pt,line cap=rect,fill=fillColor,fill opacity=0.70] (185.62, 79.70) rectangle (185.90, 89.53);
\definecolor{fillColor}{RGB}{169,169,169}

\path[draw=drawColor,line width= 0.6pt,line cap=rect,fill=fillColor,fill opacity=0.70] (194.42,167.08) rectangle (194.79,176.91);
\definecolor{fillColor}{RGB}{178,34,34}

\path[draw=drawColor,line width= 0.6pt,line cap=rect,fill=fillColor,fill opacity=0.70] (185.62,167.08) rectangle (194.42,176.91);
\definecolor{fillColor}{RGB}{169,169,169}

\path[draw=drawColor,line width= 0.6pt,line cap=rect,fill=fillColor,fill opacity=0.70] (216.13,188.92) rectangle (233.67,198.75);
\definecolor{fillColor}{RGB}{178,34,34}

\path[draw=drawColor,line width= 0.6pt,line cap=rect,fill=fillColor,fill opacity=0.70] (185.62,188.92) rectangle (216.13,198.75);

\path[draw=drawColor,line width= 0.6pt,line cap=rect,fill=fillColor,fill opacity=0.70] (185.62,112.47) rectangle (187.50,122.30);
\definecolor{fillColor}{RGB}{169,169,169}

\path[draw=drawColor,line width= 0.6pt,line cap=rect,fill=fillColor,fill opacity=0.70] (186.75,101.54) rectangle (187.00,111.37);
\definecolor{fillColor}{RGB}{178,34,34}

\path[draw=drawColor,line width= 0.6pt,line cap=rect,fill=fillColor,fill opacity=0.70] (185.62,101.54) rectangle (186.75,111.37);
\definecolor{fillColor}{RGB}{169,169,169}

\path[draw=drawColor,line width= 0.6pt,line cap=rect,fill=fillColor,fill opacity=0.70] (185.62, 90.62) rectangle (186.62,100.45);

\path[draw=drawColor,line width= 0.6pt,line cap=rect,fill=fillColor,fill opacity=0.70] (188.74,134.31) rectangle (189.07,144.14);
\definecolor{fillColor}{RGB}{178,34,34}

\path[draw=drawColor,line width= 0.6pt,line cap=rect,fill=fillColor,fill opacity=0.70] (185.62,134.31) rectangle (188.74,144.14);
\definecolor{fillColor}{RGB}{169,169,169}

\path[draw=drawColor,line width= 0.6pt,line cap=rect,fill=fillColor,fill opacity=0.70] (257.83,199.84) rectangle (347.74,209.67);
\definecolor{fillColor}{RGB}{178,34,34}

\path[draw=drawColor,line width= 0.6pt,line cap=rect,fill=fillColor,fill opacity=0.70] (185.62,199.84) rectangle (257.83,209.67);
\end{scope}
\begin{scope}
\path[clip] (  0.00,  0.00) rectangle (361.35,216.81);
\definecolor{drawColor}{gray}{0.30}

\node[text=drawColor,anchor=base east,inner sep=0pt, outer sep=0pt, scale=  0.88] at (172.57, 70.66) {Wind direction};

\node[text=drawColor,anchor=base east,inner sep=0pt, outer sep=0pt, scale=  0.88] at (172.57, 81.58) {Temperature min.};

\node[text=drawColor,anchor=base east,inner sep=0pt, outer sep=0pt, scale=  0.88] at (172.57, 92.51) {Wind direction $\times$ Region};

\node[text=drawColor,anchor=base east,inner sep=0pt, outer sep=0pt, scale=  0.88] at (172.57,103.43) {Wind mean $\times$ Region};

\node[text=drawColor,anchor=base east,inner sep=0pt, outer sep=0pt, scale=  0.88] at (172.57,114.35) {Wind direction $\times$ Wind gust max.};

\node[text=drawColor,anchor=base east,inner sep=0pt, outer sep=0pt, scale=  0.88] at (172.57,125.27) {Wind mean};

\node[text=drawColor,anchor=base east,inner sep=0pt, outer sep=0pt, scale=  0.88] at (172.57,136.19) {Wind direction $\times$ Region $\times$ Wind mean};

\node[text=drawColor,anchor=base east,inner sep=0pt, outer sep=0pt, scale=  0.88] at (172.57,147.12) {Lightning category};

\node[text=drawColor,anchor=base east,inner sep=0pt, outer sep=0pt, scale=  0.88] at (172.57,158.04) {Risk};

\node[text=drawColor,anchor=base east,inner sep=0pt, outer sep=0pt, scale=  0.88] at (172.57,168.96) {Temperature max.};

\node[text=drawColor,anchor=base east,inner sep=0pt, outer sep=0pt, scale=  0.88] at (172.57,179.88) {Wind gust max.};

\node[text=drawColor,anchor=base east,inner sep=0pt, outer sep=0pt, scale=  0.88] at (172.57,190.80) {Rain max.};

\node[text=drawColor,anchor=base east,inner sep=0pt, outer sep=0pt, scale=  0.88] at (172.57,201.73) {Region};
\end{scope}
\begin{scope}
\path[clip] (  0.00,  0.00) rectangle (361.35,216.81);
\definecolor{drawColor}{gray}{0.30}

\node[text=drawColor,anchor=base,inner sep=0pt, outer sep=0pt, scale=  0.88] at (185.62, 56.13) {0.0};

\node[text=drawColor,anchor=base,inner sep=0pt, outer sep=0pt, scale=  0.88] at (242.10, 56.13) {0.5};

\node[text=drawColor,anchor=base,inner sep=0pt, outer sep=0pt, scale=  0.88] at (298.58, 56.13) {1.0};

\node[text=drawColor,anchor=base,inner sep=0pt, outer sep=0pt, scale=  0.88] at (355.05, 56.13) {1.5};
\end{scope}
\begin{scope}
\path[clip] (  0.00,  0.00) rectangle (361.35,216.81);
\definecolor{drawColor}{RGB}{0,0,0}

\node[text=drawColor,anchor=base,inner sep=0pt, outer sep=0pt, scale=  1.10] at (266.68, 44.09) {$\Delta\ell_{\xi_0}$ in $000$'s};
\end{scope}
\begin{scope}
\path[clip] (  0.00,  0.00) rectangle (361.35,216.81);
\definecolor{drawColor}{RGB}{0,0,0}

\node[text=drawColor,anchor=base west,inner sep=0pt, outer sep=0pt, scale=  1.10] at (191.32, 14.44) {Term type};
\end{scope}
\begin{scope}
\path[clip] (  0.00,  0.00) rectangle (361.35,216.81);
\definecolor{drawColor}{RGB}{0,0,0}
\definecolor{fillColor}{RGB}{169,169,169}

\path[draw=drawColor,line width= 0.6pt,line cap=rect,fill=fillColor,fill opacity=0.70] (247.66, 11.71) rectangle (260.69, 24.74);
\end{scope}
\begin{scope}
\path[clip] (  0.00,  0.00) rectangle (361.35,216.81);
\definecolor{drawColor}{RGB}{0,0,0}
\definecolor{fillColor}{RGB}{178,34,34}

\path[draw=drawColor,line width= 0.6pt,line cap=rect,fill=fillColor,fill opacity=0.70] (294.65, 11.71) rectangle (307.68, 24.74);
\end{scope}
\begin{scope}
\path[clip] (  0.00,  0.00) rectangle (361.35,216.81);
\definecolor{drawColor}{RGB}{0,0,0}

\node[text=drawColor,anchor=base west,inner sep=0pt, outer sep=0pt, scale=  0.88] at (266.91, 15.20) {linear};
\end{scope}
\begin{scope}
\path[clip] (  0.00,  0.00) rectangle (361.35,216.81);
\definecolor{drawColor}{RGB}{0,0,0}

\node[text=drawColor,anchor=base west,inner sep=0pt, outer sep=0pt, scale=  0.88] at (313.89, 15.20) {smooth};
\end{scope}
\end{tikzpicture}
    \caption{Total in-bag log-likelihood increase for each of the predictors until the stopping iteration has been reached -- sums up to $\approx2400$. The plot refers to the short forecast model. Smooth base learners included: P-splines, Ordinal smooths, and Markov Random Fields.}
    \label{fig:varimp_12_bern}
\end{figure}

Sixty-two percent of all observations were zeros, so the likelihood of the binary model  was calculated from almost three times as many observations as that of positive continuous models (see equation \ref{eq:zadj_proof}), thus potentially making the parameter $\xi_0$ the most influential one. Figure \ref{fig:varimp_12_bern} depicts, which variables contributed the most to the increases in log-likelihood after the offset term was set. Region was by far the most important predictor and was composed of two base learners of similar importance: the ridge-penalised contrasts for independent between-region differences and the MRF smoother for neighbourhood-based spatial differences. Second, the \textit{Wind gust maximum} variable was highly important on its own, but it did not interact with \textit{Region} and \textit{Wind direction} as much as the \textit{Mean wind}. On average, the model predicted zero faults with $\hat{p}\approx69\%$ for zero observations and $\hat{p}\approx50\%$ for nonzero observations, which suggests a higher leverage of zero counts.

For the continuous case, the results of the four initial models are outlined in the top four rows of table \ref{table:pos_cont}. Based on these initial models, it was likely that a four-parameter distribution was optimal, so two more distributions of that type were trialled. The Generalised Beta 2 is presented in the fifth row of table \ref{table:bern}, and it severely underperformed. The Box-Cox Power Exponential Distribution, on the other hand, performed much better but converged extremely slowly and had to be stopped after making only minor improvements from the $>7000$'th iteration onwards with $\ell_{total}> -500$.

\begin{table}
\centering
\begin{tabular}{l*{4}{c}}
\hline
$\mathcal{D}$ & $K$ & $\ell_{total}$ & $m_{stop}$ & $\ell_{average}$ \\
\hline 
GA   & 2 & -1244 & 1545 & -0.61 \\ 
GIG  & 3 & -839  & 1315 & -0.41 \\ 
GG   & 3 & -626  & 1494 & -0.30 \\ 
BCTo & 4 & -189  & 5146 & -0.09 \\
GB2  & 4 & -6297 & 4700 & -2.29 \\
\hline
\end{tabular}
\caption{Positive continuous models' out-of-sample performance. Distributions' acronyms stand for Gamma, Generalised Inverse Gaussian, Generalised Gamma, Box-Cox t, Generalised Beta 2 distribution (in order).}
\label{table:pos_cont}
\end{table}

The Box-Cox \textit{t}-distribution performed by far the best out of all models and converged to a solution within a reasonable training time. The same distribution was, therefore, used to train a gradient-boosted CIT model. Hyperparameter tuning was done using the binary genetic algorithm described in section \ref{sec:hyper_details}. After two days of parallel computing on eight-CPU Windows PC with Intel i7 9700k processor, the model yielded out-of-sample $max(\ell_{total}^{(i)}) = \{-376, -330, -330\}$ for generations $i \in [1..3]$. The best result was obtained using \textit{maximum depth} = 2, \textit{number of sampled columns} = 7, and \textit{minimum number of observations on a leaf} = 450. 

However, the tree model was still significantly worse than the GAM model, so the latter was selected and refitted to all nonzero training observations. The resulting model had an in-bag performance of $\ell_{total} = 494$ and $\ell_{avg} \approx 0.061$. The in-bag performance was beyond excellent, but it likely resulted from overfitting of the model, which is much more acceptable in the domain of purely predictive models.

On testing data, the binary model had $\ell_{total} = -2906.8$ and $\ell_{avg} \approx -0.557$, which was better than its performance on the validation data (see table \ref{table:bern}). The positive continuous model also improved with $\ell_{total} \approx -29.2$ and $\ell_{avg} \approx -0.0145$ compared to its results in table \ref{table:pos_cont}. 

\begin{figure}
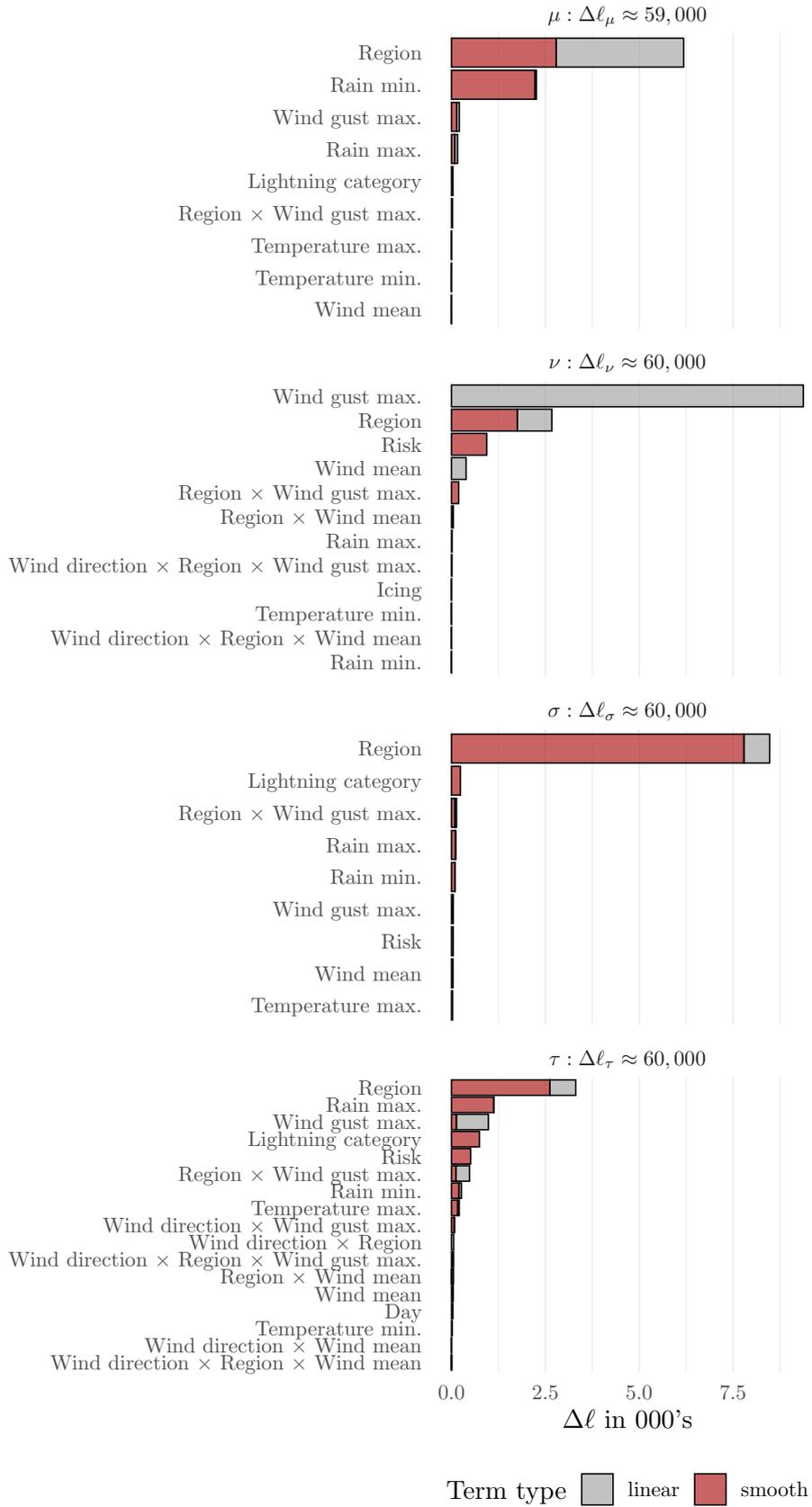

    \centering
    \include{figures/bcto_varimp_plot}
    \caption{Analogous to figure \ref{fig:varimp_12_bern} except that the model was only trained on positive counts -- $\approx38\%$ of all observations. Notice the x-axis scale differed between the two figures. The intercept term was removed for legibility reasons, but its improvements were included in the $\Delta\ell_{\theta_k}$ estimates on top.}
    \label{fig:varimp_12_pos_cont}
\end{figure}

The variable importance plots for each of the parameters are presented in figure \ref{fig:varimp_12_pos_cont}. Overall, the estimates of all parameters were much more accurate thanks to the predictor variables, which further encouraged the use of a four-parameter LSS model. In comparison, the Bernoulli model benefitted little from the data as its $\Delta\ell_{\xi0}\approx2,400$ despite using three times as many observations for that effect. In both models, the spatial effects made a significant contribution to the overall likelihood. The scale and kurtosis parameters ($\sigma$ and $\tau$) benefitted especially strongly from spatial neighbourhood effects between the regions.

\begin{figure}
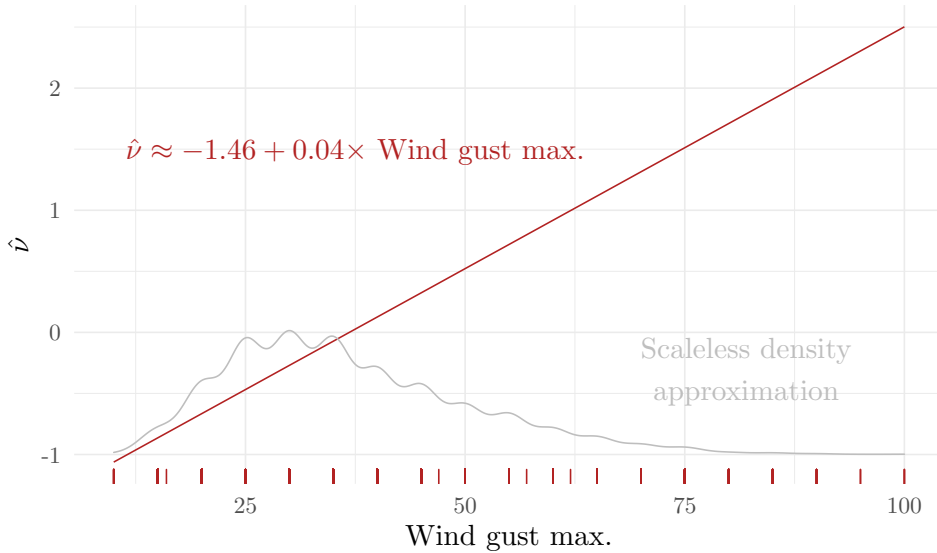

    \centering
    \include{figures/partial_nu_wind_gust_1_2}
    \caption{Partial effect plot of \textit{Wind gust maximum} on the skewness parameter $\nu$.}
    \label{fig:partial_nu}
\end{figure}

Interestingly, the best predictor of the skewness parameter $\nu$ was, by far, the linear predictor of \textit{Wind gust maximum}. Figure \ref{fig:partial_nu} shows the effects of changes in wind gust on the skewness of the distribution assuming all other variables were held constant and without accounting for the interactions with \textit{Region}. \textit{Wind gust max.} was associated with right skew for all values below 36.5 and with left skew for all values above that threshold.

\begin{figure}
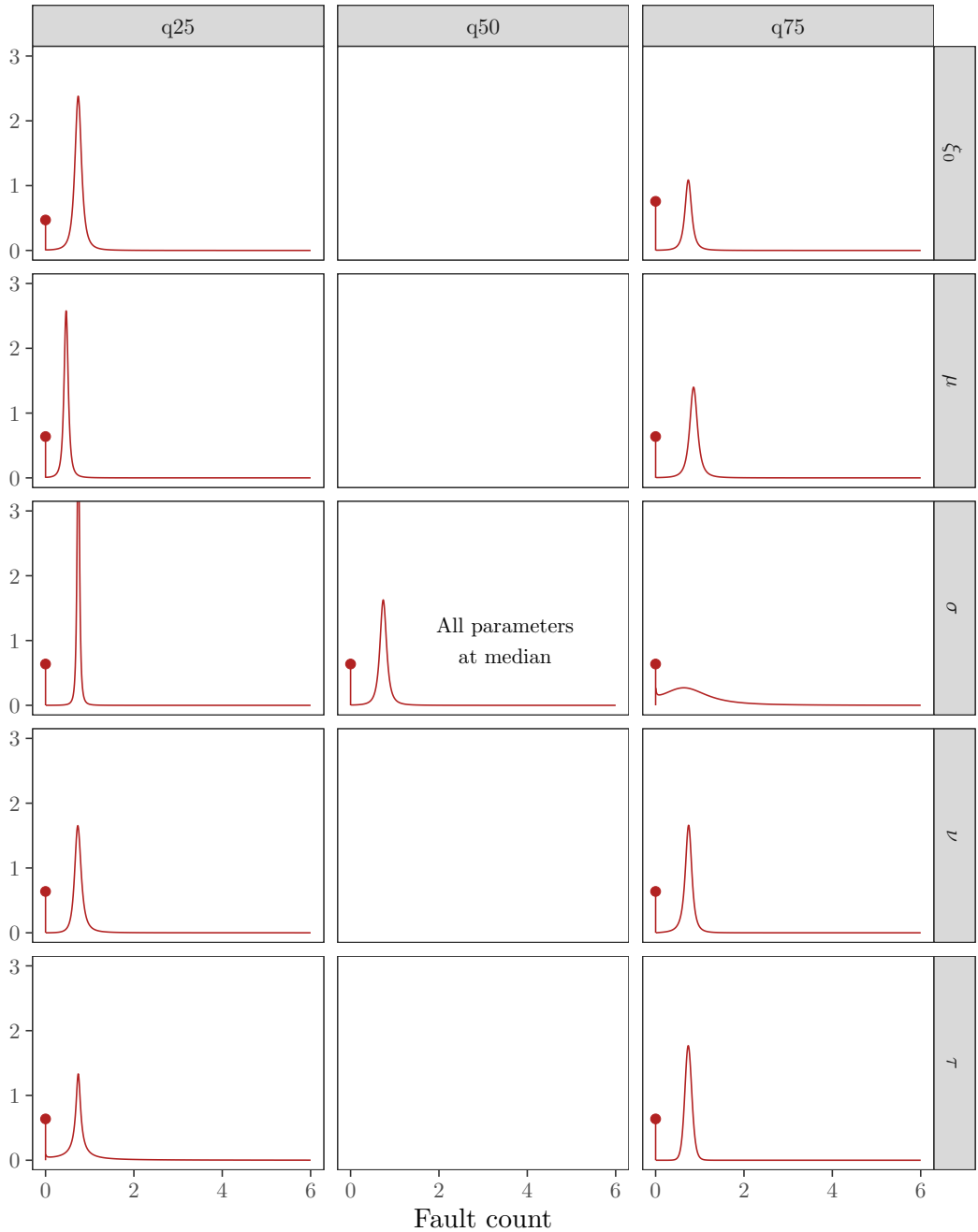

    \centering
    \include{figures/zadj_bcto_vis}
    \caption{Zero-adjusted Box-Cox t distributions based on quantiles of the fitted parameter values on the testing data. Each parameter is, by default, held at its median value. Plots in the 'q25' and 'q75' columns have one parameter (indicated on the right) changed to its first or third quartile respectively (indicated by q25 or q75 respectively). The red dot indicates the probability mass of the observation being equal to zero.}
    \label{fig:zadj_bcto_vis}
\end{figure}

\subsubsection{Probability of surpassing a fault threshold} \label{sec:prob_above_thresh}

\begin{figure}
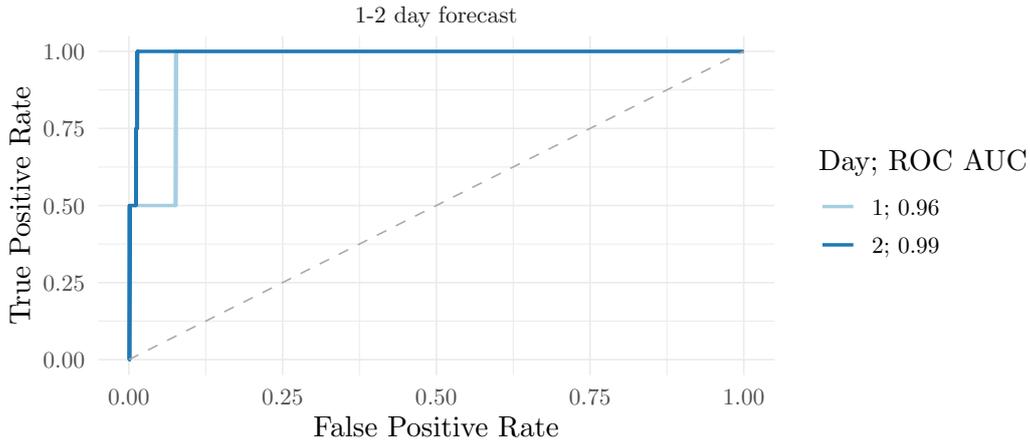

    \centering
    \include{figures/roc_curves}
    \caption{ROC curves for the probability of faults being greater or equal to 14 using the testing data.}
    \label{fig:roc_curves}
\end{figure}

The electricity company has also mentioned that it is of special interest to accurately predict the probability of surpassing a certain number of faults, because the penalty severity depends heavily on it. One such threshold was the threshold of 14 faults or more per region. Figure \ref{fig:roc_curves} shows two Receiver Operating Characteristic (ROC) curves  based on the cumulative density function of the predicted zero-adjusted distribution and the testing data. High area under the curve (AUC) scores suggest that the model was well suited for predicting the probability of faults surpassing the threshold of 14.

\section{Discussion}

This study shows a novel approach for probabilistic forecasting of overhead distribution lines' failures based on weather data. Our zero-adjusted Box-Cox \textit{t} model provides a consistent method for agile data-driven decision making  without the need of altering observations' weights during model fitting to compensate for varying error costs. Using gradient boosting (GB) allowed for efficient and automatic variable and term selection as well as overall performance improvements. Surprisingly, the GB Generalised Additive Models (GB GAMs) surpassed the widely popular and extremely effective tree-based methods, such as XGBoost and GB Conditional Inference Trees (CIT). In GB GAMs, the fault region was the single most important predictor of every parameter except for skewness. Much of its predictive performance came from the spatial structure of the regions and the information about their neighbouring areas.

\subsection{Extending previous methods}

Earlier machine learning approaches were forced to optimise complex multi-objective functions at the fitting stage in order to satisfy metrics based on misclassification rates \citeaffixed{Kabir2019}{e.g.,}. Instead, our approach focused solely on predicting the distribution of faults with the aim of maximising the out-of-sample log likelihood. This metric is rarely used explicitly in machine learning but often implicitly through out-of sample root mean squared error, which yields equivalent models to those selected using out-of-sample Gaussian likelihood. 

When the model appropriately predicts the distribution of faults, the decisions can be made more flexibly based on the available organisational resources and the most up-to-date penalty policy. Other methods \citeaffixed{Kabir2019}{e.g.,} require changing the weights of the observations or resampling to account for asymmetric misclassification costs.

It is unclear how decisions should be made when the model consistently outputs biased probabilities (e.g., by artificially overestimating the baseline probability of observing nonzero faults). Some decision-makers might try to selectively over- or underestimate the threat of faults to account for changing organisational resources. This decision-making process would end up being highly opinionated, erratic, and hence prone to human error. 

Our method, on the other hand, provides the decision-makers with access to one information only -- the most likely distribution of the number of faults. They can then use that information to decide themselves how to minimise potential losses. This approach places a higher burden on the decision-makers but should ultimately lead to lower costs than relying on a model to have all of the nuance already built into its inner workings. 

We have also shown that using a smoothed generalised additive structure can surpass the frequently used black-box tree-based models. The likely reason for this finding is that prior research only compared standard semi-parametric GAMs to tree-based models using ensemble methods, such as bagging or boosting \cite{Kabir2019}, even though these methods can also be applied to improve the performance of GAMs \cite{Mayr2014ExtendingDevelopments}. Among other benefits, GAMs offer more natural ways for modelling ordinal, cyclic, and spatial information. Furthermore, GB GAMs offer superb diagnostic capabilities due to their simple additive structure.

\subsection{Presenting forecasts to decision-makers}

One of the major cons of this model is that it might be rather unintuitive to nonstatistical audiences. Its optimal use requires a solid understanding of uncertainty, probability density functions, and integrals. Some of these requirements can be lowered by changing the presentation of the forecasts. There should also be a way to determine whether each new weather forecast is sufficiently similar to the ones used for the training of the model. Without this safeguard, the model could quietly extrapolate and create nonsensical predictions.

\subsubsection{Summarising the distribution}

Business professionals often demand discretisation of information, because many decisions require nothing more than a few summary statistics. Demands are placed for point estimates representing the worst-case, most plausible, and best-case scenarios. From a practical perspective, our model can only provide the worst-case scenario without misleading its users. First, the best-case scenario would (almost) always equal zero, given that the model predicted $\xi_0 > 0.14$ for all testing observations. Second, zero-inflated distributions are bimodal, which means that any single estimate of central tendency would be highly misleading. The worst-case scenario can be interpreted as a really high percentile of the predicted distribution (e.g., the 99th) and is within our model's reach.



\subsubsection{Extrapolation warning}

The model might sometimes come across weather forecasts which are highly dissimilar to any weather forecasts from the training data. If, for example, the predicted maximum temperature is higher than it has ever been, then the model should inform the decision-maker of that fact, and perhaps show what the forecast would be if the temperature was within the previously observed range. Our model was not designed for extrapolation, so extreme caution and domain expertise need to be utilised in such circumstances.



Incidentally, our model used no smooth interactions, so multivariate relationships could easily be investigated by combining the variables together into linear interactions. For example, a linear interaction of maximum temperature and wind mean could be examined by simply multiplying the two variables together, and then checking whether the new observation lies within the range of the training data. A three-way interaction of wind mean (continuous), wind direction (eight categories), and region (nine or three categories) could be checked by (1) subsetting the training data such that all observations have the same wind direction and region as the new observation, and (2) checking if the new observation's mean wind speed falls within the range of the subsetted training data.

\subsection{Limitations and future research}

As with most machine learning applications, the models could have been improved by enhancing data quality, conducting a finer hyperparameter search, and/or advancing computational and methodological aspects of the utilised models.



\subsubsection{Data}

An obvious data limitation was that the regions from the forecast provider did not overlap perfectly with the company's historical administrative regions. This led to the need for aggregating fault counts by area-based weighted averaging, thus leading to inaccurate count estimates. A count model would restrict the availability of distributions for tuning purposes, but the increased precision would likely compensate for these limitations.

Also, the data included weather predictions over multiple time intervals, which were then aggregated into a single observation for modelling. We could assume that the partial effect of weather at each interval is exactly the same conditioning on all other weather variables. Aggregation allows for imposing such constraint on linear base learners given that

\begin{align} 
    \begin{split}\label{eq:agg_interval_values}
    \hat{\bm{u}} &= \dots+\beta_x\bm{x}^{(1)} +
    \beta_x\bm{x}^{(2)}+\dots+\beta_x\bm{x}^{(T)} \\
    &= \dots+\beta_x(\bm{x}^{(1)}+\bm{x}^{(2)}+\dots+\bm{x}^{(T)}) \\ 
    &\propto\dots+\beta_x\hat{\bm{x}}
    \end{split}
\end{align}

where $T$ is the number of time intervals over which the variable $\{\bm{x}^{(1)},\ldots,\bm{x}^{(T)}\}$ was recorded. However, for smooth terms this equation does not work. For example, if two extreme temperatures are observed on the same day, they might cause more faults than observing two average temperatures. At the same time, using $T$ separate terms could lead to an overparametrised model, so it would be best to impose the restriction on all terms to have the same partial effect. 

Because P-splines are based on a penalised linear equation, the same method from equation \ref{eq:agg_interval_values} could be adopted for them. Before fitting, all $T$ measurements for $\bm{x}$ would need to be transformed into a B-Spline basis with the same knot locations and degree. A new basis matrix would need to be defined by aggregating the bases' values at the same knots (columns) for the same observations (rows) across all bases matrices for each of the time intervals: $\bm{B}^{(new)} := \sum_{t=1}^T(\bm{B}^{(t)}) \propto \big\{\sum_{t=1}^T(\bm{B}^{(t)})\big\}/T$. The penalty and smoothing matrices would stay exactly the same. 

Another limitation was that the forecaster's wind direction variable was binned into eight categories instead of supplying the data as a continuous angle variable, which unnecessarily simplified the data. Furthermore, the angle variable was supplied as an arithmetic mean over the four six-hour intervals, which did not allow for aggregating the records by summing their bases -- analogously to the examples from the previous paragraph.


\subsubsection{Methodological improvements}

The models described in this study were based on statistical developments, which are still in their early stages at the time of writing. One of them would be to allow for the inclusion of a cyclic base learner with one degree of freedom in order to ensure unbiased term selection as suggested by \citeasnoun{Hofner2011ABoosting}. The package \texttt{mboost} does not allow for it at the time of writing as it uses \possessivecite{Fahrmeir2004PenalizedPerspective} method, which requires inverting the cyclic difference matrix, which is singular. 

\paragraph{Subsampling at each iteration} \label{sec:subsampling}

Another potentially useful feature would be the implementation of random subsampling of observations before each boosting iteration. This idea was first introduced by \citeasnoun{Friedman2002StochasticBoosting} who coined the term \textit{Stochastic Gradient Boosting} for it. The smaller the proportion of the subsampled rows, the higher the regularisation of each individual learner and its variance. However, the base learners end up being less correlated with each other, which is likely the cause for an overall variance reduction of the full aggregated model.

Stochastic Gradient Boosting is often used with tree models (e.g., XGBoost), so it would be a welcome addition to the \textit{gamboostLSS}' GB CITs. It might also work well for GB GAMs, but this has not yet been tested at the time of writing. In the worst case, different subsampling rates would be trialled, and the rate of one would be selected based on its superior performance.

Ensemble tree models also often use random subsampling of variables, which includes all tree models from our study (i.e. XGBoost and CIT). This is another potential opportunity for improving GB GAMs as it also induces regularisation, reduces the correlation between learners, and speeds up computation time. 

Both of these subsampling methods have the potential to improve performance but can also make it more difficult to interpret the outputs and variable importance plots as we have done in section \ref{sec:res_short_forecasting}. It is virtually impossible to take into account which observations and variables were selected at each iteration, so using a stochastic algorithm could make it more difficult to establish under what conditions the different base learners were selected. Another problem for GB GAMs is to decide whether they should subsample variables or base learners.

\paragraph{Accounting for parameter uncertainty}

Another problem with our models was that they neglected the uncertainty around the parameter estimates. The predictions would be based on a distribution $\mathcal{D}(\hat{\xi}_0, \hat{\bm{\theta}})$, but its parameters were all calculated by a deterministic model estimated from the data. Uncertainty around model estimates could be achieved by switching to Bayesian methods \cite{Umlauf2018BAMLSS:Beyond}, but that would come at a large computational cost. Model selection would need to be handled using a LASSO-type penalty \cite{Groll2019LASSO-typeShape} -- which is comparable in many respects to early stopping in gradient boosting \cite{Hepp2016ApproachesLasso}.

There are also some alternatives for the estimation of parameter uncertainty under standard gradient boosting methods. One of them is bootstrapping, but the biased nature of gradient boosting would make the inferences invalid. Instead, semi-parametric GAMs often utilise the Bayesian interpretation of the penalised likelihood methods in order to run a posterior simulation on the base learners \cite[p. 293]{Wood2017GeneralizedR}. For a smooth base learner with a basis matrix $\bm{B}$, we could take the estimates $\bm{\beta}$ and their covariance matrix $\Sigma$. Then simulate new $\bm{\beta}^{(i)}$ vectors from the multivariate normal distribution, \textit{MVN}$(\bm{\beta}, \Sigma)$. Next, multiply each of these vectors by the basis matrix $\bm{B}$ and sum its rows to obtain the posterior predictions. This method assumes that the smoothing parameter $\lambda$ is fixed but bootstrapping it makes little difference.

\paragraph{Other ideas}

\citeasnoun{Duan2020NGBoost:Prediction} have developed another alternative to cyclic and noncyclic updating of location, scale, and shape models. In their method, all parameters are regressed simultaneously at each iteration; however, they do not use standard gradients as the response but their transformation -- the natural gradients. Normally, standard gradients differ vastly between different parameters of the same distribution, because they operate in the distribution space, where the distances between distributions are unstable and dependent upon individual parameterisations instead of the objective loss function (negative log likelihood in this case). By multiplying the standard gradients by the inverse of the Fisher's information, the distances between parameters in the distribution space start to correspond to the distances of the parameters in the parameter space (as in standard gradient descent). This method holds a lot of potential but is still in its early stages of development and has not yet been applied to the methods used in this study.

\subsubsection{Computational improvements}

Model-based boosting still has space for improvement in terms of computational efficiency. Faster model convergence not only saves time, but also allows for trying out more model settings, which then lead to better performance. 

Beyond the possibility of subsampling described in section \ref{sec:subsampling}, the packages lacked parallel computing support within individual models. Every iteration, all base learners were fitted to the gradients of the loss function w.r.t. each of the parameters. All of them were fitted independently of each other, so parallelising this process should be a natural extension to this method. Lack of GPU support has also put our methods at a disadvantage compared to the extremely efficient \textit{machine learning} packages, such as \texttt{xgboost} or \texttt{LightGBM}.

Moreover, there have been significant computational developments in terms of training individual base learners more efficiently. \possessivecite[p. 290]{Wood2017GeneralizedR} \texttt{mgcv} package has the function \texttt{bam()} for fitting optimised generalised additive models for big data. \citeasnoun{Li2020FasterCovariates} created an algorithm, which drastically reduces computation time of the matrix cross-product $\bm{X}^\intercal\bm{WX}$, which is a major computational bottleneck in GAMs. In the one-dimensional case, the algorithm discretises the $n$ rows of the $n \times p$ matrix $\bm{X}$, such that it can be transformed into an $m \times p$ matrix $\overline{\bm{X}}$ with $m$ distinct rows. By doing that, the algorithm avoids the biggest computational bottleneck without any noticeable changes to the model fit. The function also allows for parallel computation and takes additional care to speed up the computation of the Cholesky decomposition of the previously mentioned matrix cross-product. 

Lastly, the hyperparameter tuning of the XGBoost model was limited by the \texttt{tidymodels} package, which does not support the tuning of the cost complexity parameter $\gamma$, shrinkage parameter $\lambda$, and the mixing parameter $\alpha$.

\section{Conclusion}

This study showed a new paradigm for supporting management of overhead distribution lines via probabilistic predictive modelling using gradient-boosted models for location, scale, and shape. Smooth and linear base learners proved that they can surpass the performance of the commonly used tree models in this field.


\newpage
\bibliography{references.bib}
\bibliographystyle{agsm}
\newpage
\includepdf[pages=-, angle=91,scale=.7,pagecommand=\section*{Appendix A}\label{Appendix_A}]{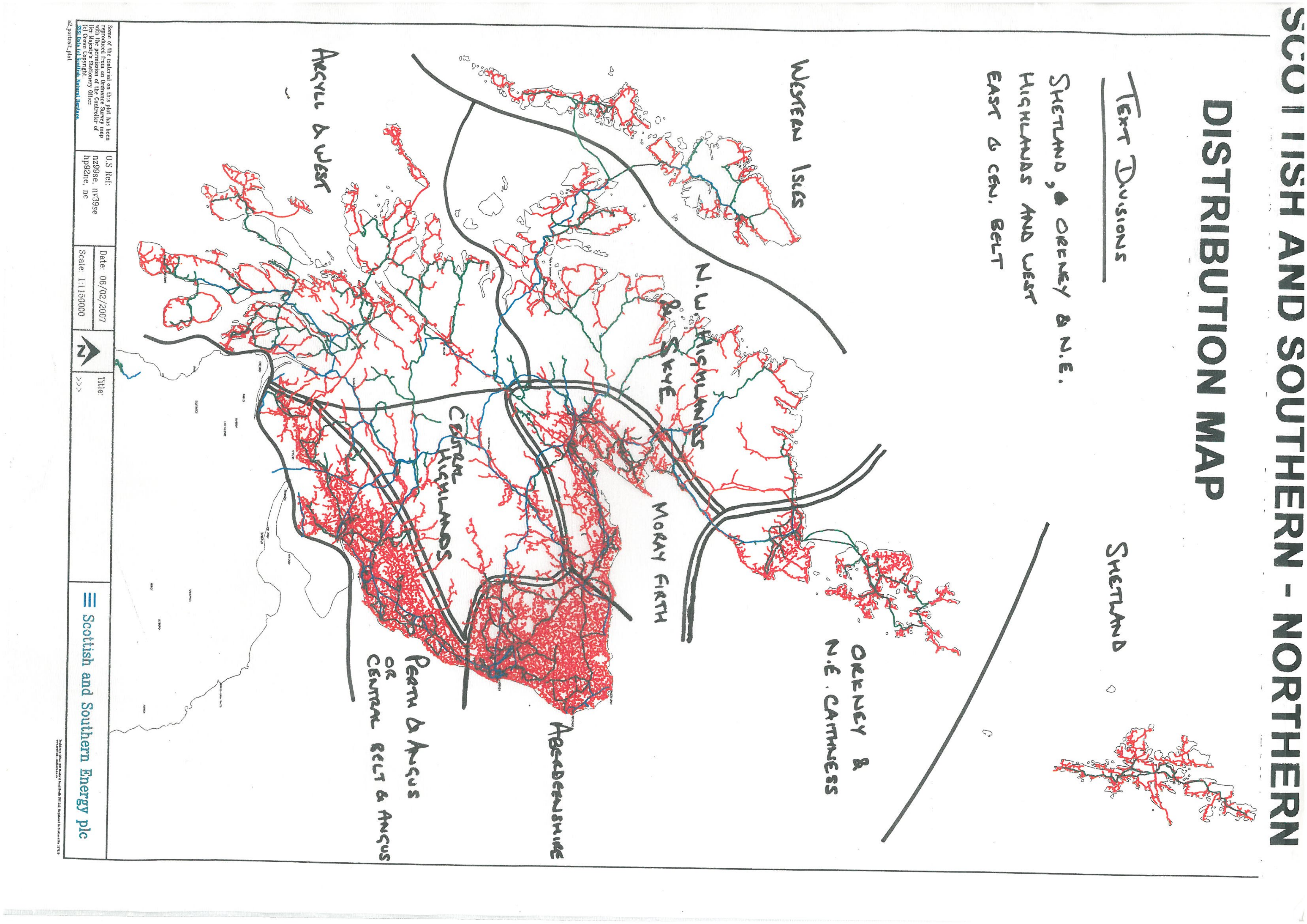}

\section*{Supplementary materials}

All code used to produce the results is available on \href{https://github.com/asieminski/odl-fault-pred}{GitHub}.
\end{document}